\documentclass[preprint,pre,12pt]{revtex4-1}
\usepackage{amsfonts}
\usepackage{amssymb}
\usepackage{amsmath}
\usepackage{graphicx}

\begin{document}
\title{Calculation of mean spectral density for statistically uniform   tree-like random models}
\author{E. Bogomolny and O. Giraud}
\affiliation{Univ.~Paris-Sud, CNRS, LPTMS, UMR8626, F-91405, Orsay, France}
\date{October 04, 2013}

\begin{abstract}

For random matrices with tree-like structure there exists a recursive relation for the local Green functions whose solution permits to find directly many important quantities in the limit of infinite matrix dimensions. The purpose of this note is to investigate and compare  expressions for the spectral density of random regular graphs, based on  easy approximations for real solutions of the recursive relation valid for trees with large coordination number. The obtained formulas  are in a good agreement with the results of numerical calculations even for small coordination number.
 
\end{abstract}

\maketitle


\section{Introduction}\label{introduction}

Matrices with random (or pseudo-random) elements appear naturally in many different problems and are well investigated in physical and mathematical literature (see e.g.~\cite{mehta, oxford}). In this note we consider a special class of sparse random matrices, namely, matrices associated with tree (or tree-like) structures. A fundamental property of such matrices  is that the number of non-zero elements in each row and column  either remains finite or grows much slowly than the matrix dimension when the latter increases.   

As usual, a (connected) tree is a graph where any pair of vertices is connected by only one path without repeating vertices. Vertices are labeled by integers. If a symmetric (or Hermitian) matrix $M$ is such that its entries $M_{ij}$, $i\neq j$, are non-zero if and only if vertices $i$ and $j$ are connected on a given tree, then the matrix $M$ is said to be associated with the tree. An example of such a matrix is the adjacency matrix of the tree. In general, diagonal entries of $M$ are nonzero.

Let $G_{ij}$ be the Green function for matrix $M$, namely
\begin{equation}
G_{mn}(E)=\left (M-E\ I\right )^{-1}_{mn }\ ,
\label{green}
\end{equation}
where $I$ is the identity matrix.   
It is well established that for matrices associated with a tree  there exist  recursive relations  which  connect the diagonal elements $G_{ii}$ of the Green function with similar quantities but for smaller matrices. Probably the simplest way to derive  such  relations on a tree is to use an easily verified identity (sometimes called the Schur complement formula), which states that for any matrix $M$ one has
\begin{equation}
G_{mm}(E)=\left (M_{mm}-E -\sum_{k,p} M_{mk}\tilde{G}_{kp}^{(m)}(E) M_{pm} \right )^{-1}\ ,
\label{general}
\end{equation}
where  $\tilde{G}_{kp}^{(m)}(E)$ is the Green function as in \eqref{green} but for the matrix  $\tilde{M}^{(m)}$ obtained from $M$ by removing row $m$ and column $m$. 

For a tree, removing one vertex splits the remaining graph into a disjoint union of smaller trees. Thus, $\tilde{M}^{(m)}$ is block-diagonal and the Green function $\tilde{G}_{kp}^{(m)}(E)$  has no matrix elements between different neighbors of a fixed site $m$. It means that for trees  Eq.~\eqref{general} takes the form 
\begin{equation}
G_{mm}(E)=\left (M_{mm}-E -\sum_k |M_{mk}|^2\tilde{G}_{kk}^{(m)}(E) \right )^{-1}\ ,
\label{Gmm}
\end{equation}
where the sum is taken over all neighbors of site $m$ and we assume that matrix $M$ is Hermitian.

Applying the same arguments to  $\tilde{G}_{kp}^{(m)}(E)$ leads to a similar equation for each neighbor $k$ of $m$, namely 
\begin{equation}
\tilde{G}_{kk}^{(m)}(E)=\left (M_{kk}-E -\sum_{p\neq m} |M_{kp}|^2\tilde{G}_{pp}^{(k)}(E) \right )^{-1}\ ,
\label{tree_equation}
\end{equation}
where the sum is over all neighbors of $k$ except $m$ (which has already been removed). In principle, $\tilde{G}_{pp}^{(k)}(E)$ in the right-hand side of this equation is the Green function element for a matrix obtained from $M$ by removing two connected sites $m$ and $k$. But because we are on a tree, $\tilde{G}_{pp}^{(k)}(E)$ is the same whether only $k$ or all its ancestors are removed; thus it is sufficient to indicate only the last removed site. It is this property which permits to write the recursive relation \eqref{tree_equation} where on both sides similar quantities are present. 

For finite trees the above relations allow to calculate  the Green function recursively, but their the most important application corresponds to infinite (or very large) trees where matrix elements of $M_{mn}$ are assumed to be independent random variables (or constants).  In this case, because of the disjoint nature of different sub-trees, Eq.~\eqref{tree_equation} for all $k$ only involves variables $M_{kp}$ from the sub-tree to which $k$ belongs. Thus for large uniform trees, one can assume that all $\tilde{G}_{kk}^{(m)}$ are independent random variables having the same distribution. Let $K+1$ be the coordination number of vertex $k$, and $G_p$, $1\leq p\leq K$, be random variables distributed according to that distribution. For simplicity we assume below that the diagonal matrix element $M_{kk}$  is a random variable $e$, off-diagonal elements $M_{kp}$ are real i.i.d.~variables $V_p$, and all coordination numbers are equal to $K+1$.  Then  Eq.~\eqref{tree_equation} means  that the random variable
\begin{equation}
G=\frac{1}{e -E-\sum_{p=1}^{K} V_p^2 G_p}
\label{te}
\end{equation} 
has the same distribution as the $G_p$. 

This equation is the main tool for the investigation of random uniform trees. It has been obtained initially by Abou-Chacra,  Thouless, and  Anderson in their study of self-consistent theory of localization \cite{bethe_tree} and later it has been re-derived by many different methods: replica formalism   \cite{rodgers},  $\sigma$-model \cite{fyodorov}, Ricatti equation \cite{derrida}, rank-one perturbation \cite{flambaum}, cavity method \cite{mezard} etc. We shall refer below  to \eqref{te} as the tree equation. 

Strictly speaking, the tree equation \eqref{te} is valid only for infinite uniform trees (the Bethe lattice) but in many cases it is applied to models which have only tree-like structure, i.e.~which can locally be approximated by trees but may have loops of large length. A typical example is that of random regular graphs, where each vertex has the same number of neighbors, as in a tree with constant coordination number, but where the boundary shell present in finite trees is absent (see e.g.~\cite{mezard_parisi}). For certain tree-like models  the validity of the tree equation \eqref{te} can be  proved rigorously  \cite{resolvent,geisinger}.  

There exist three main types of tree-like problems, corresponding to three possible sources of randomness in the tree equation. The first corresponds to a regular tree with a fixed coordination number and with only diagonal disorder (i.e.~$e$ is a random variable and $V_p=1$). This model has been proposed in a seminal paper \cite{bethe_tree} and was recently investigated in \cite{biroli_1} and \cite{biroli_2}. The second class (see e.g. \cite{off-diagonal}) corresponds to models which are also defined on a fixed regular tree but have only off-diagonal disorder (i.e. all  $V_p$  are i.i.d.~random variables and $e=0$). Finally, the third  type of models includes trees without disorder but with fluctuating coordination number. Characteristic examples of such models are Erd\"os-R\'enyi graphs \cite{regular_graph} or  sparse random matrices with a finite connectivity \cite{rodgers}. In this note, as an example, we will restrict ourselves to random regular graphs.

A  general method for numerically solving the tree equation \eqref{te} has been proposed in  \cite{bethe_tree} and it is commonly used now under the name of belief propagation method \cite{mezard}. The main steps of this method are as follows. First, fix arbitrarily an initial sample of a large number, say $N$, of elements $G_k$, $k=1,\ldots, N$. Second, choose randomly $K$ integers from $1$ to $N$ and variables $e$ and $V_p$ from their known distributions. Third, calculate $G$ from the tree equation \eqref{te}. Fourth, choose randomly an element from the initial sample and replace it by the calculated $G$. Repeat these steps till the convergence of the resulting distribution to the distribution of $G$. There exist two types of solutions of \eqref{te}. The first one corresponds to real values of the energy $E$: at each step of the iteration, the variables $G_k$ are real and after iteration one obtains the distribution of the real variable $G$. The second type of solution corresponds to adding a small positive imaginary part to the energy, that is, put $E\to E+i\eta$ in \eqref{te}. In this case, upon iteration the $G_k$ become complex variables and the result of the iteration will yield the distribution of both the real and the imaginary part of $G$, which may and will depend on $\eta$. If the energy $E$ corresponds to a region of localized states, then when $\eta\to 0$ the imaginary part of $G$ will tend to zero almost everywhere, while if the energy corresponds to a region of non-localized states the imaginary part of $G$ goes to a finite distribution. Here we restrict ourselves to real values of the energy, with $\eta=0$.

The purpose of this note is to discuss the calculation of the mean spectral density of typical tree-like models by construction of approximate solution of the tree equation. Strictly   speaking, the method is  valid for trees with a large coordination number but often gives good results even at small coordination numbers. The method itself is not new and has been used widely for random  Erd\"os-R\'enyi graphs (see e.g.~\cite{network}--\cite{semerjian} and references therein). We first clarify certain important points which seem not to be discussed in the literature and then  investigate in detail the application of the method to random regular graphs and compare different types of approximations. 

The plan of the paper is the following. After setting some definitions in Section \ref{rhodeE}, Section~\ref{mean_field} is devoted to the general discussion of the method. It is demonstrated that the commonly used 'mean-field' solution of the tree equation corresponds to the approximation of the exact solution by a symmetric Cauchy distribution whose parameters are calculated self-consistently. Different useful formulas for the mean spectral density are briefly discussed in this Section.  In Section~\ref{diagonal} the case  of regular graphs with diagonal disorder is considered. By comparing results of direct numerical calculations with various kinds of approximations we check their precision  and found that at zeroth order the best results for regular trees with diagonal disorder is given by the so-called 'single defect approximation' proposed in \cite{monasson}.  Section~\ref{off_diagonal} treats the case of regular graphs with off-diagonal disorder. For these models the best results are obtained by using the modified effective medium approximation introduced in Section~\ref{mean_field}.   In all considered cases, the next order approximation for the tree equation solution, though it agrees much better with the numerical solution obtained by belief propagation, improves noticeably the spectral density only at lowest coordination numbers. The conclusion of this note (stated in Section~\ref{conclusion}) is that the approximate solution of the tree equation is useful, flexible, and general method of calculation of the mean spectral density in various uniform tree-like models.           


\section{Spectral density}\label{rhodeE}

The mean spectral density  is defined as usual by 
\begin{equation}
\label{defrho}
\rho(E)=\left \langle \frac{1}{N}\sum_i \delta(E-E_i)\right \rangle =\frac{1}{N\pi}\sum_n \left \langle \mathrm{Im}\ G_{nn}(E)\right \rangle\ ,
\end{equation}
where $E_i$ are eigenvalues of $M$ and the average is performed over random realizations of the matrix entries. 

As mentioned in the introduction, here we only consider real solutions of Eq.~\eqref{te}. Let us denote by $r(e)$ and $p(V)$ the (known) probability densities of $e$ and $V$. The probability density of variable $G\equiv x$ is given by some function $g(x)$, and that  of variable $ z=\sum_{j=1}^K V_j^2x_j$ by a function $F_K(z)$. By definition, $F_k(z)$ is
\begin{equation}
F_K(z)=\int \delta \left (z-\sum_{j=1}^K V_j^2x_j \right ) \prod_{j=1}^K p(V_j)g(x_j)\, \mathrm{d}V_j\, \mathrm{d}x_j\ .
\label{sum_k}
\end{equation}
The tree equation \eqref{te} implies that for real $x$ the probability density $g(x)$  satisfies  the equation
\begin{equation}
g(x)=\int \delta\left (x-\frac{1}{e-E-z}\right ) r(e)\, F_K(z)\, \mathrm{d}e\, \mathrm{d}z\ ,
\label{general_g}
\end{equation}
or equivalently
\begin{equation}
g(x)=\frac{1}{x^2}
\int F_K\left (e-E-\frac{1}{x} \right )r(e)\, \mathrm{d}e\ .
\label{general_g2}
\end{equation}
Eigenvalues of $M$ correspond to values of $E$ where $G_{mm}(E)$ gets singular. Assuming that $M_{mm}$ and $\tilde{G}_{kk}^{(m)}(E)$ in \eqref{Gmm} can be replaced by random variables, which we denote by $e$ and $x_k$ respectively, we see that these singularities occur at $E=e-\sum_{j=1}^{K+1} V_j^2x_j$. Under these assumptions, the definition \eqref{defrho} gives
\begin{equation}
\rho(E)=\int \delta \left ( e-E-\sum_{j=1}^{K+1} V_j^2x_j \right )r(e)\, \mathrm{d}e\prod_{j=1}^{K+1} p(V_j)g(x_j)\, \mathrm{d}V_j\, \mathrm{d}x_j=\int F_{K+1}(e-E)r(e)\mathrm{d}e\ ,
\label{general_rho}
\end{equation} 
where the summation is performed over all $K+1$ neighbors of a given site. 

Another useful expression for the mean spectral density is obtained by transforming Eq.~\eqref{general_g} into the form
\begin{equation}
\frac{1}{y^2}g\left (\frac{1}{y}\right )=\int \delta \left (y-\Big (e-E-\sum_{j=1}^{K} V_j^2x_j \Big )\right ) r(e)\,\mathrm{d}e\prod_{j=1}^{K} p(V_j)g(x_j)\, \mathrm{d}V_j\, \mathrm{d}x_j\ .
\end{equation}
Performing the integral over $x_{K+1}$ in Eq.~\eqref{general_rho} yields
\begin{equation}
\rho(E)=\int \frac{1}{y^2 V^2 }g\Big (\frac{1}{y}\Big )g\Big (\frac{y}{V^2}\Big )p(V)\,  \mathrm{d}V\, \mathrm{d}y\ .
\label{modified_rho}
\end{equation}
Of course, the above formulas can be rewritten in many equivalent forms.

Therefore the knowledge of the spectral density can directly be deduced from that of the distribution $g(x)$ which is a solution of the tree equation \eqref{te}, or from that of $F_K(z)$. In the next section we consider simple solutions of these equations.


\section{'Mean-field'  approximation and beyond}\label{mean_field}

\subsection{Constant solution and single defect approximation}
The tree equation \eqref{te} may have a constant complex solution, that is, a $\tilde{x}$ (depending on $E$) such that
\begin{equation}
\tilde{x}=\int \Big (e-E -\tilde{x}\sum_{j=1}^K V_j^2\Big )^{-1}r(e)\, \mathrm{d}e\prod_{j=1}^K p(V_j)\,  \mathrm{d}V_j \ . 
\label{ema}
\end{equation}
If such a $\tilde{x}=\tilde{x}(E)$ exists, the mean spectral density associated with this solution is obtained from \eqref{general_rho} as
\begin{equation}
\rho(E)=\int \delta \left ( e-E-\tilde{x}(E)\sum_{j=1}^{K+1} V_j^2\right )r(e)\, \mathrm{d}e\prod_{j=1}^{K+1} p(V_j)g(x_j)\, \mathrm{d}V_j\, \mathrm{d}x_j\ .
\label{general_rhobis}
\end{equation} 
Using contour integration, this expression can be equivalently rewritten as
\begin{equation}
\rho_{\mathrm{SDA}}(E)=\frac{1}{\pi}\, \mathrm{Im}\, \int \Big (e-E -\tilde{x}(E)\sum_{j=1}^{K+1} V_j^2\Big )^{-1}r(e)\, \mathrm{d}e\prod_{j=1}^{K+1} p(V_j)\,  \mathrm{d}V_j\ 
\label{rho_sda}
\end{equation}
when the imaginary part of $\tilde{x}(E)$ is positive. Equation \eqref{rho_sda} is known as the single defect approximation (SDA) \cite{monasson}. 

To further simplify this expression one can simply argue that for large $K$ (which, as we see below, is the parameter which control this type of approximation) the sum over $K+1$ terms in Eq.~\eqref{rho_sda} can be approximated by a sum over $K$ terms, so that from \eqref{ema} one concludes that the mean spectral density takes the form
\begin{equation}
\rho_{\mathrm{EMA}}(E)=\frac{1}{\pi}\, \mathrm{Im} \, \tilde{x}(E) .
\label{rho_ema}
\end{equation}
This type of approximation is widely applied for sparse random matrices, where Eq.~\eqref{rho_ema} is called the effective medium approximation (EMA) \cite{network,semerjian}. Physically, this solution is a kind of a mean field  and often these equations are obtained by arguing that the random variable $x$ distributed according to $g(x)$ fluctuates slowly around its mean value $\tilde{x}$ (see e.g. \cite{network}).  

Such an approach is simple, physically transparent, does not require heavy numerical calculations, and gives, as a rule, quite good results. The trouble is that its main assumption that variable $x$ fluctuates only slowly around its mean value $\tilde{x}$ cannot be, in general, correct for tree-like models. Indeed, it follows from Eq.~\eqref{general_g2} that if the function $F_K(x)$ is smooth, then $g(x)$ has to decrease as
\begin{equation}
g(x)\underset{x\to\infty}{\sim}\frac{C}{x^2},\qquad C=\int F_K(e-E)r(e)\, \mathrm{d}e\ .
\label{limit}
\end{equation}
In particular, this means that $g(x)$ belongs to the class of heavy-tail distributions, for which the mean value $\int_{-\infty}^{\infty}xg(x)\mathrm{d}x$ does not exist.  Therefore, the meaning of  a (complex) mean field solution \eqref{ema} and, especially, its relation to a direct numerical  solution of the tree equation \eqref{te} for real $x$ remains obscure.

\subsection{Cauchy distribution and  the modified effective medium approximation}
The above mean field approach is equivalent to the assumption that the function $g(x)$ can be approximated by the symmetric Cauchy distribution
\begin{equation}
g_{\xi}(x)=\frac{\gamma}{\pi(\gamma^2+(x-\mu)^2)}=\frac{1}{\pi}\mathrm{Im}\frac{1}{x-\xi}
\label{cauchy}
\end{equation}     
with $\mu$ and $\gamma$ real parameters, $\gamma>0$, and $\xi=\mu+\mathrm{i}\gamma$. Indeed, for a test function $\varphi(x)$ without singularities in the upper-half plane one has the identity
\begin{equation}
\label{cauchydelta}
\int_{-\infty}^{\infty}\varphi(x) g_{\xi}(x)\mathrm{d}x=\varphi(\xi)=\int_{-\infty}^{\infty}\varphi(x) \delta(x-\xi)\mathrm{d}x\ .
\end{equation}
In other words, the Cauchy distribution \eqref{cauchy} is indistinguishable from the $\delta$-function when acting on a large class of functions. In particular, when $a$ and $b$ are real and  $b>0$, contour integration yields the following formula
\begin{equation}
\int_{-\infty}^{\infty}\exp \left (\frac{\mathrm{i} \lambda}{a-b x}\right )\frac{\gamma}{\pi(\gamma^2+(x-\mu)^2)}\mathrm{d}x=
\exp \left (\frac{\mathrm{i}\lambda}{a-b(\mu+\mathrm{i}\, \mathrm{sign} (\lambda) \gamma)}\right )\ .
\label{useful_int}
\end{equation}
Other useful elementary properties of Cauchy distributions that we will make use of are the following identities
\begin{equation}
\label{idcauchy1}
\frac{1}{z^2}g_{\xi}\left(\frac{1}{z}\right)=g_{1/\xi^*}(z)\ ,\qquad
\lambda g_{\xi}(\lambda z)=g_{\xi/\lambda}(z)\ ,
\end{equation}
and the convolution property
\begin{equation}
\label{idcauchy2}
\int_{-\infty}^{\infty}g_{\xi_1}(x)g_{\xi_2}(z-x)\mathrm{d}x=g_{\xi_1+\xi_2}(z)\ .
\end{equation}
The characteristic function of \eqref{cauchy} is given by
\begin{equation}
\label{characteristicCauchy}
\hat{g}_{\xi}(\lambda)= \mathrm{e}^{\mathrm{i}\lambda \mu -|\lambda|\gamma} \ .
\end{equation}
Let us calculate the characteristic function of $g(x)$,
\begin{equation}
\hat{g}(\lambda)=\mathbb{E}(\mathrm{e}^{\mathrm{i}\lambda x})\equiv \int_{-\infty}^{\infty} \mathrm{e}^{\mathrm{i}\lambda x}g(x)\mathrm{d}x\ .
\end{equation}    
Using  \eqref{general_g} one obtains the exact functional relation
\begin{equation}
\label{gandg}
\hat{g}(\lambda)=\int \exp \left (\frac{\mathrm{i} \lambda}{e-E-\sum_{j=1}^K V_j^2 x_j}\right  )r(e)\mathrm{d} e\prod_{j=1}^K g(x_j)\mathrm{d}x_jp(V_j)\,  \mathrm{d}V_j \ .
\end{equation} 
If in the right-hand side of this equation $g(x)$ is replaced by the Cauchy distribution $g_{\xi}(x)$ given by Eq.~\eqref{cauchy}, the consecutive use of \eqref{useful_int}  gives a ''first-order approximation'' of the characteristic function of $g$ as
\begin{equation}
\hat{g}^{(1)}(\lambda)=\int \exp \left (\frac{\mathrm{i} \lambda}{e-E-(\mu +\mathrm{i}\, \mathrm{sign} (\lambda) \gamma) \sum_{j=1}^K V_j^2  }\right )r(e)\mathrm{d} e \prod_{j=1}^K p(V_j)\,  \mathrm{d}V_j\ .
\label{hat_g_1}
\end{equation} 
In general $\hat{g}^{(1)}(\lambda)\neq \hat{g}_{\xi}(\lambda)$. Requiring that parameters $\mu$ and $\gamma$ be such that at small argument
\begin{equation}
\lim_{\lambda\to 0} \ln \hat{g}_{\xi}(\lambda) =\lim_{\lambda\to 0} \ln \hat{g}^{(1)}(\lambda)
\label{consistency}
\end{equation}
is exactly equivalent to requiring that $\tilde{x}=\xi$ satisfies the 'mean field' equation \eqref{ema}.

These arguments can be reformulated as follows. Let $\tilde{x}$ be a (complex) constant solution of \eqref{ema} and $g_{\tilde{x}}(x)$ the corresponding Cauchy distribution. Then an approximate solution to \eqref{general_g} such that $\hat{g}(\lambda)=\hat{g}_{\tilde{x}}(\lambda)$ for small arguments $\lambda\to 0$ is obtained by replacing $g$ by $g_{\tilde{x}}$ in the right-hand side of Eq.~\eqref{gandg}.

It means that $\tilde{x}$ is not a mean value of a random variable as in the usual mean field approach but just a  pole of the Cauchy distribution which reproduces the behavior of the characteristic function $\hat{g}(\lambda)$ at small $\lambda$. (It is interesting to notice that for the usual GOE ensemble of random matrices the local Green function also has a Cauchy distribution, see e.g.~\cite{yan}.)

The knowledge of $\tilde{x}$ permits to calculate easily the mean spectral density from Eqs.~\eqref{rho_sda} or \eqref{rho_ema}.
In the cases considered in the next Sections we found that approximation \eqref{rho_ema} does not give good results at small values of $K$. Instead, we propose to use another simple approximation obtained by substituting the Cauchy approximation \eqref{cauchy} into  Eq.~\eqref{modified_rho}. Simple transformations based on Eqs.~\eqref{idcauchy1}--\eqref{idcauchy2} lead to the following approximate formula for the mean spectral density
\begin{equation}
\rho_{\mathrm{MEMA}}(E)=-\frac{1}{\pi}\mathrm{Im}\int  \frac{1}{\tilde{x}y^2-\tilde{x}^{-1}}P(y)\mathrm{d}y\ .
\label{mema}
\end{equation}
This expression is almost as simple as EMA \eqref{rho_ema} but often gives better results. We refer to it as to the modified effective medium approximation (MEMA).

\subsection{First order approximations}
A few  formulas are useful to mention. When $g_{\tilde{x}}(x)$ is substituted into Eq.~\eqref{sum_k} one gets, after integration over $x_j$, a first order approximation of $F_K(z)$, as
\begin{equation}
F_K^{(1)}(z)=\frac{1}{\pi} \mathrm{Im}\int \frac{1}{z-\tilde{x}\sum_{j=1}^KV_j^2}\prod_{j}p(V_j)\, \mathrm{d}V_j,\, \qquad \tilde{x}=\mu+\mathrm{i}\gamma \ .
\label{f_k_cauchy}
\end{equation}
The first iteration of the initial Cauchy distribution, $g^{(1)}(x)$, is then straightforwardly obtained either from this expression and Eq.~\eqref{general_g}, or directly from Eq.~\eqref{hat_g_1}: if we define 
\begin{equation}
\tilde{\mathcal{X}}_K=\frac{1}{e-E-\tilde{x}\sum_{j=1}^KV_j^2}\ ,
\end{equation} 
then Eq.~\eqref{hat_g_1} can be rewritten
\begin{equation}
\hat{g}^{(1)}(\lambda)=\left\{\begin{array}{cc}
\langle \exp(i\lambda\tilde{\mathcal{X}}_K)\rangle&\,\,\lambda>0\\
\langle \exp(i\lambda\tilde{\mathcal{X}}_K^*)\rangle&\,\,\lambda<0
\end{array}
\right.
\end{equation}
where average is taken over $e$ and the $V_j$. This is exactly the characteristic function of the Cauchy distribution with parameters $(\textrm{Re}\tilde{\mathcal{X}}_K,\textrm{Im}\tilde{\mathcal{X}}_K)$ (see Eq.~\eqref{characteristicCauchy}). Thus one directly gets 
\begin{equation}
g^{(1)}(x)=\langle g_{\tilde{\mathcal{X}}_K}(x)\rangle=\frac{1}{\pi} \mathrm{Im}\int  \frac{1}{x-\tilde{\mathcal{X}}_K}\,  r(e)\, \mathrm{d}e\prod_{j=1}^K p(V_j)\,  \mathrm{d}V_j\ .
\label{g_1}
\end{equation}  
The next iterations can be calculated in the same manner.

Below we refer to  the Cauchy function $g_{\xi}(x)$ with parameters from Eq.~\eqref{ema} as the zeroth order approximation of $g(x)$ and to the first iteration \eqref{g_1} as the first order approximation. For each order of approximation one can calculate the mean spectral density by using either Eq.~\eqref{general_rho} or Eq.~\eqref{modified_rho}. Both formulas are exact when $g(x)$ obeys the tree equation \eqref{te}, but for approximate expressions they may and will give different results. Though, in principle, one can rely on the results only when these two expressions are close to each other, for a given problem usually one of these formulas works better and we shall indicate it.       

 
\section{Diagonal disorder}\label{diagonal}

Let us now consider in detail the example of a regular tree (which will be approximated by a regular graph to avoid boundary effects) where each vertex has $K+1$ neighbors. The matrix we consider is the adjacency matrix of the graph, to which we add a diagonal part given by i.i.d. random variables with e.g. the uniform distribution
\begin{equation}
r(e)=\left \{\begin{array}{cc}\dfrac{1}{w},&|e|\leq \dfrac{w}{2}\\ 0,& \mathrm{otherwise.}  \end{array}\right.
\label{uniform}
\end{equation}   
It is this model which has been considered in  \cite{bethe_tree} and later has been investigated in many places (see e.g. \cite{biroli_1, biroli_2} and references therein). 

Such a model corresponds to the situation discussed in Section \ref{mean_field} with fixed $V_j=1$. Therefore the 'mean field' equation \eqref{ema} reads
\begin{equation}
\tilde{x}=\int \Big (e-E-\tilde{x} K \Big )^{-1}r(e)\, \mathrm{d}e\ .  
\label{ema_diagonal}
\end{equation}

Equation \eqref{sum_k} implies that $F_K$ is the the probability distribution of a sum of $K$ i.i.d.~variables distributed according to the law $g(x)$. From \eqref{general_g} it follows that $g(x)$ is determined by the $K$-fold convolution of $g(x)$ (cf.~\eqref{sum_k}).   When $K$ is large, the generalized central limit theorem (see e.g. \cite{limit_distribution, stable}) can be applied. It states that a properly normalized sum of i.i.d. random variables tends to the one of known stable distributions. If the second moment is finite, the limiting distribution is the Gaussian. For heavy-tail distributions, when the second moment does not exist, the limiting distribution is one of the Levy distributions. The asymptotics \eqref{limit} means that $g(x)$ belongs to the domain of attraction of the symmetric Cauchy distribution, so  at large $K$ the probability density of $K$-copies of $x$ in the bulk is close to 
\begin{equation}
\label{approxfk}
F_K(x)\underset{K\to\infty}{\sim}\frac{\Gamma}{\pi(\Gamma^2+(x-\mathcal{M})^2)}=\frac{1}{\pi}\mathrm{Im}\frac{1}{x-K\tilde{x}}\ ,
\end{equation} 
characterized by the two parameters $\Gamma=K\gamma$ and $\mathcal{M}=K\mu$, with $\tilde{x}=\mu+\mathrm{i}\gamma$.  Important is that these parameters are determined only from  the behavior of the characteristic function of $g(x)$ at small $\lambda$,
\begin{equation}
\lim_{\lambda\to 0} \ln \hat{g}(\lambda)=-\gamma |\lambda|+\mathrm{i}\lambda \mu
\end{equation}
as   in \eqref{consistency}. 

The mean spectral density is then given by \eqref{general_rho} which, using the approximation \eqref{approxfk}, yields
\begin{equation}
\rho(E)=\frac{1}{\pi}\mathrm{Im}\int\frac{1}{e-E-(K+1)\tilde{x}}r(e)\, \mathrm{d}e\ ,
\label{final_rho}
\end{equation}   
which corresponds exactly to SDA \eqref{rho_sda}. Note that the function $F_{K+1}(e_j-E)$ has the meaning of a strength function \cite{wigner} where one uses as initial wave function the one localized at a site with energy $e_j$. 

The simplest case of the above formalism corresponds to the absence of disorder, i.e. $r(e)=\delta(e)$. In this case Eq.~\eqref{ema_diagonal} gives, for $|E|\leq 2\sqrt{K}$, a solution $\tilde{x}=\mu+\mathrm{i}\gamma$ with
\begin{equation}
\mu=-\frac{E}{2K},\qquad \gamma=\frac{1}{K}\sqrt{K-\frac{E^2}{4}},
\label{shape}
\end{equation}
while  for $|E|\geq 2\sqrt{K}$, $\tilde{x}=-E/(2K)$. From Eq.~\eqref{final_rho}, the mean spectral density in this case is given at order 0 by
\begin{equation}
\rho^{(0)}(E)=\frac{K+1}{2\pi((K+1)^2-E^2)}\sqrt{4K-E^2},\qquad |E|\leq 2\sqrt{K}\ ,
\label{mckay}
\end{equation}
which agrees with Kesten-McKay law for random regular graphs \cite{kesten, regular_graph}. 

\begin{figure}

\begin{center}
\includegraphics[width=.7\linewidth,clip]{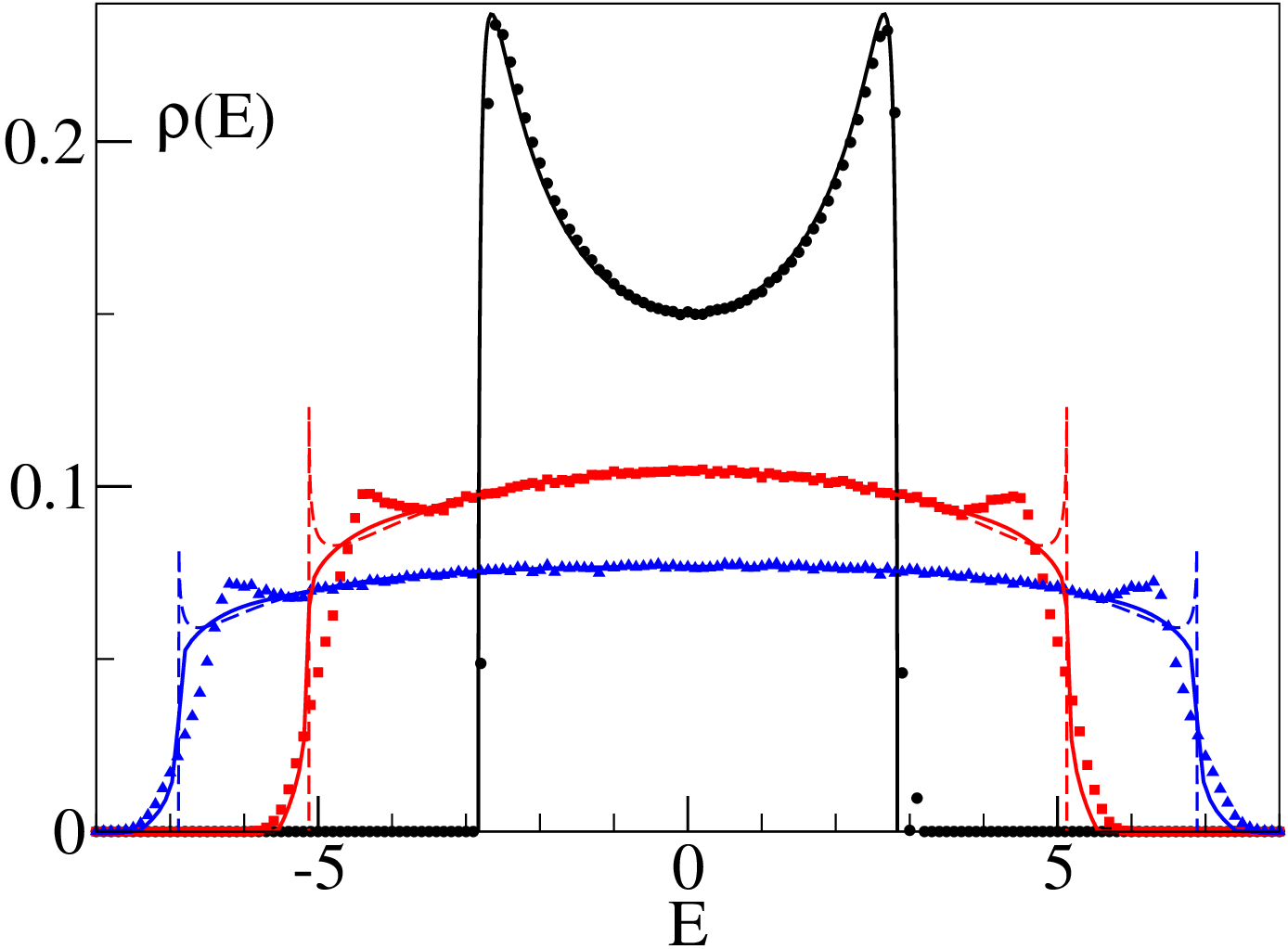}
\end{center} 

\caption{Mean spectral density for regular graphs with uniform diagonal disorder \eqref{uniform} with $K=2$ and  $W=0.3$ (black circles), $W=8$ (red squares), and $W=12$ (blue triangles), obtained from computations done on $1000$-vertex regular graphs, with $2000$ different realizations of graphs and disorder.  Dashed lines of corresponding color indicate zeroth order SDA formulas \eqref{final_rho}. Solid lines corresponds to  the next  iteration of the density \eqref{next_rho}. For $w=0.3$ these two curves are practically the same and only SDA approximation is shown.}
\label{fig_diag_0.3}
\end{figure}

\begin{figure}
\begin{minipage}{.49\linewidth}
\begin{center}
\includegraphics[width=.96\linewidth,clip]{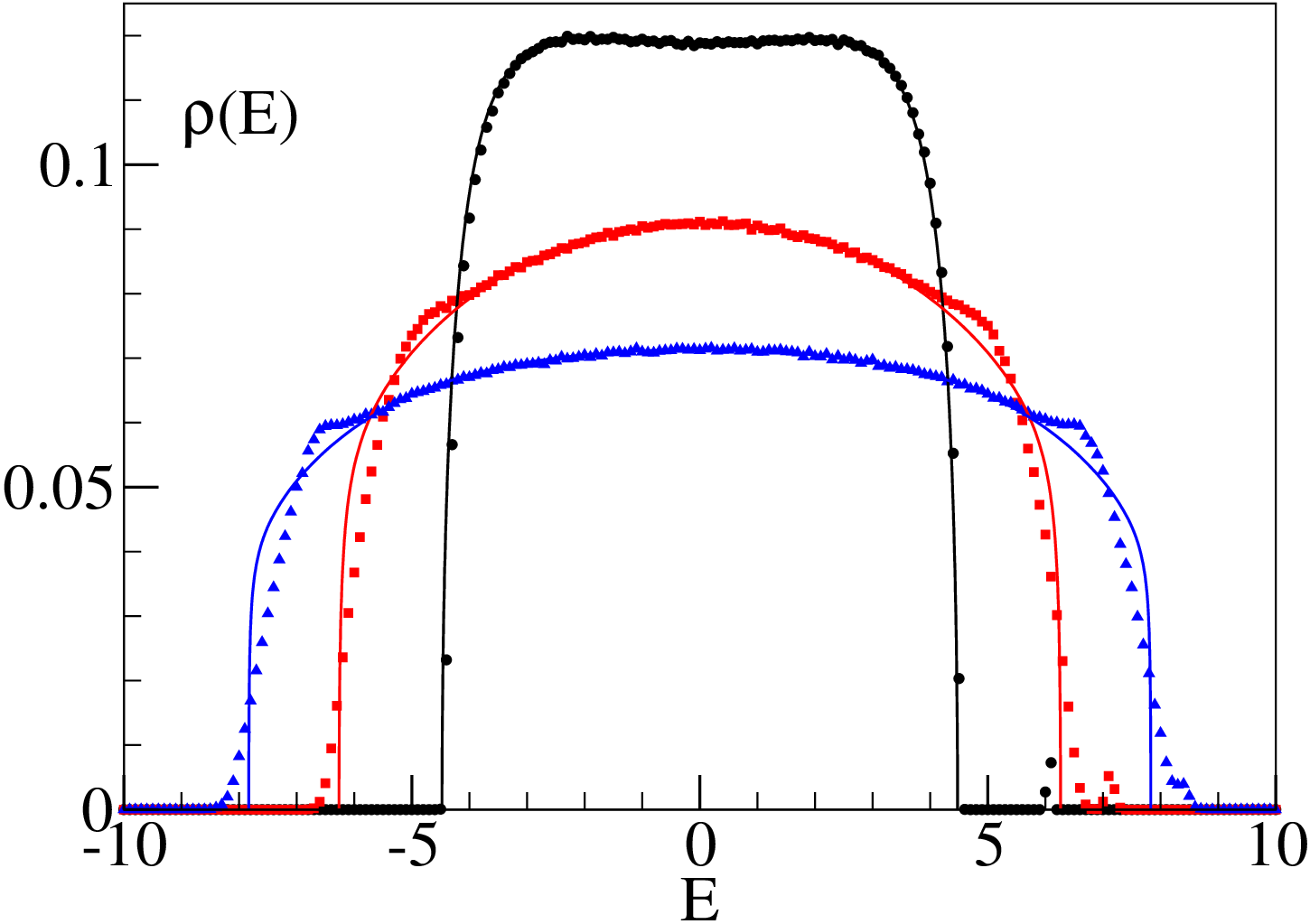}
\end{center}
\end{minipage}
\begin{minipage}{.49\linewidth}
\begin{center}
\includegraphics[width=.96\linewidth,clip]{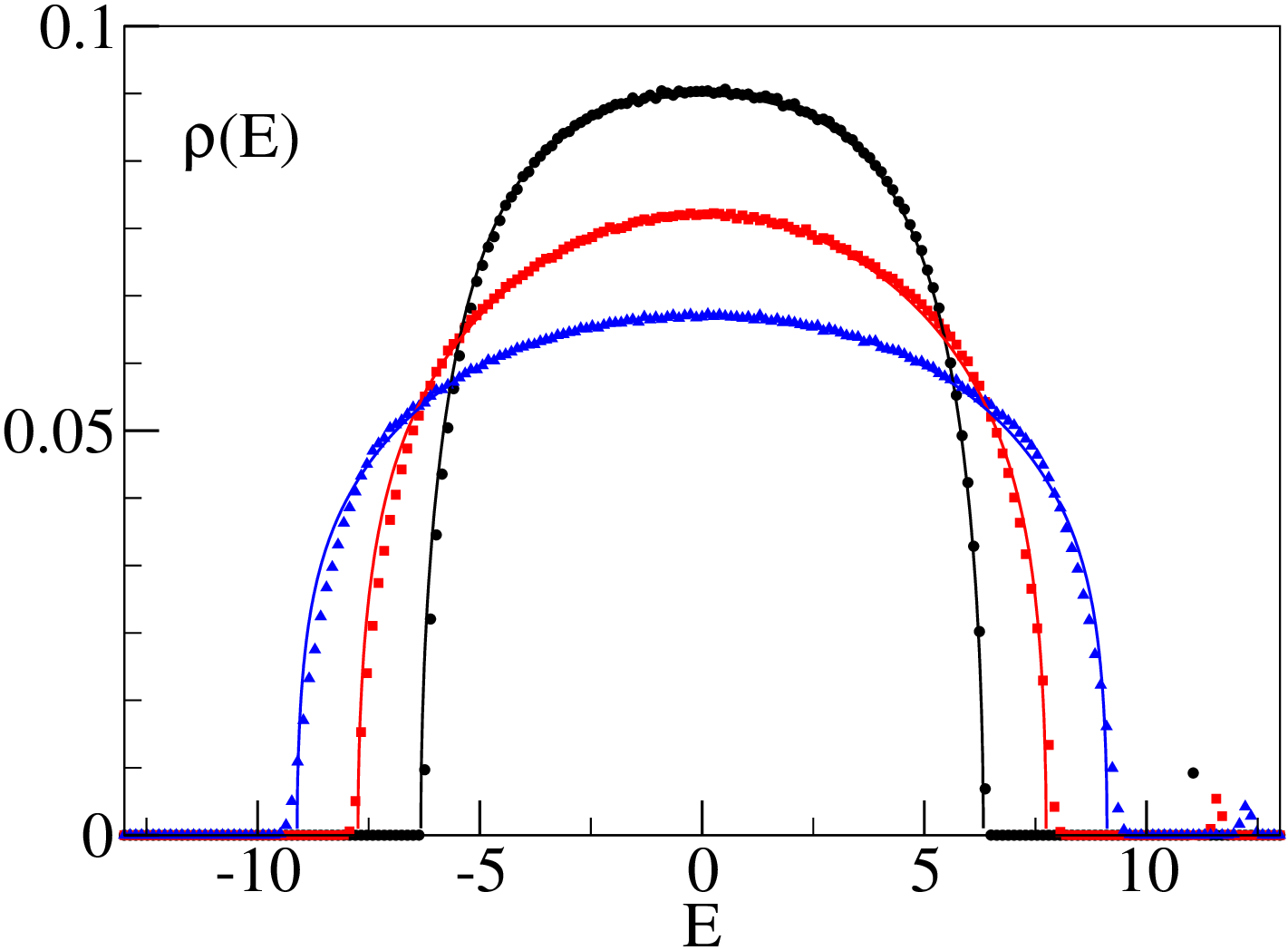}
\end{center}
\end{minipage}
\caption{Same as in Fig.~\ref{fig_diag_0.3} but for $K=5$ (left panel) and $K=10$ (right panel). Solid lines of corresponding color indicate zeroth order SDA formulas \eqref{final_rho}. The next order approximations at the scale of the figures are hardly distinguishable from the SDA and are not presented. }
\label{fig_graph}
\end{figure}

For a general distribution $r(e)$ of diagonal elements, such as \eqref{uniform}, parameters $\gamma$ and $\mu$ are easily calculated numerically from Eq.~\eqref{ema_diagonal}. Qualitatively, they have a shape (as a function of $E$) similar to \eqref{shape}. A simple way to find $\tilde{x}$ is the direct iteration of Eq.~\eqref{ema_diagonal} starting from a complex initial guess $\tilde{x}_0$, i.e. $\tilde{x}_{n+1}=f(\tilde{x}_n)$ where $f(\tilde{x})$ denotes the rhs of Eq.~\eqref{ema_diagonal}. To ensure stability it is convenient to use a slightly different iteration scheme,  $\tilde{x}_{n+1}=\alpha f(\tilde{x}_n)+(1-\alpha)\tilde{x}_n$ with a certain $\alpha<1$. Once parameters $\gamma$ and $\mu$ are tabulated as functions of energy, the spectral density can be calculated either from Eq.~\eqref{modified_rho} or from Eq.~\eqref{final_rho}. We found that Eq.~\eqref{final_rho} gives better results at small $K$.  The SDA \eqref{final_rho} is plotted for a few values of $K$ and $W$ in Figs.~\ref{fig_diag_0.3} and \ref{fig_graph} for the uniform disorder distribution \eqref{uniform}. The agreement between the above simple formulas and  the results of direct numerical diagonalization of the corresponding matrices is quite good in the bulk of the spectra even for the smallest value of the coordination number. Deviations are clearly present only near the spectral ends, where in all cases the discussed approach cannot be applied.  

\begin{figure}
\begin{center}
\includegraphics[width=.7\linewidth,clip]{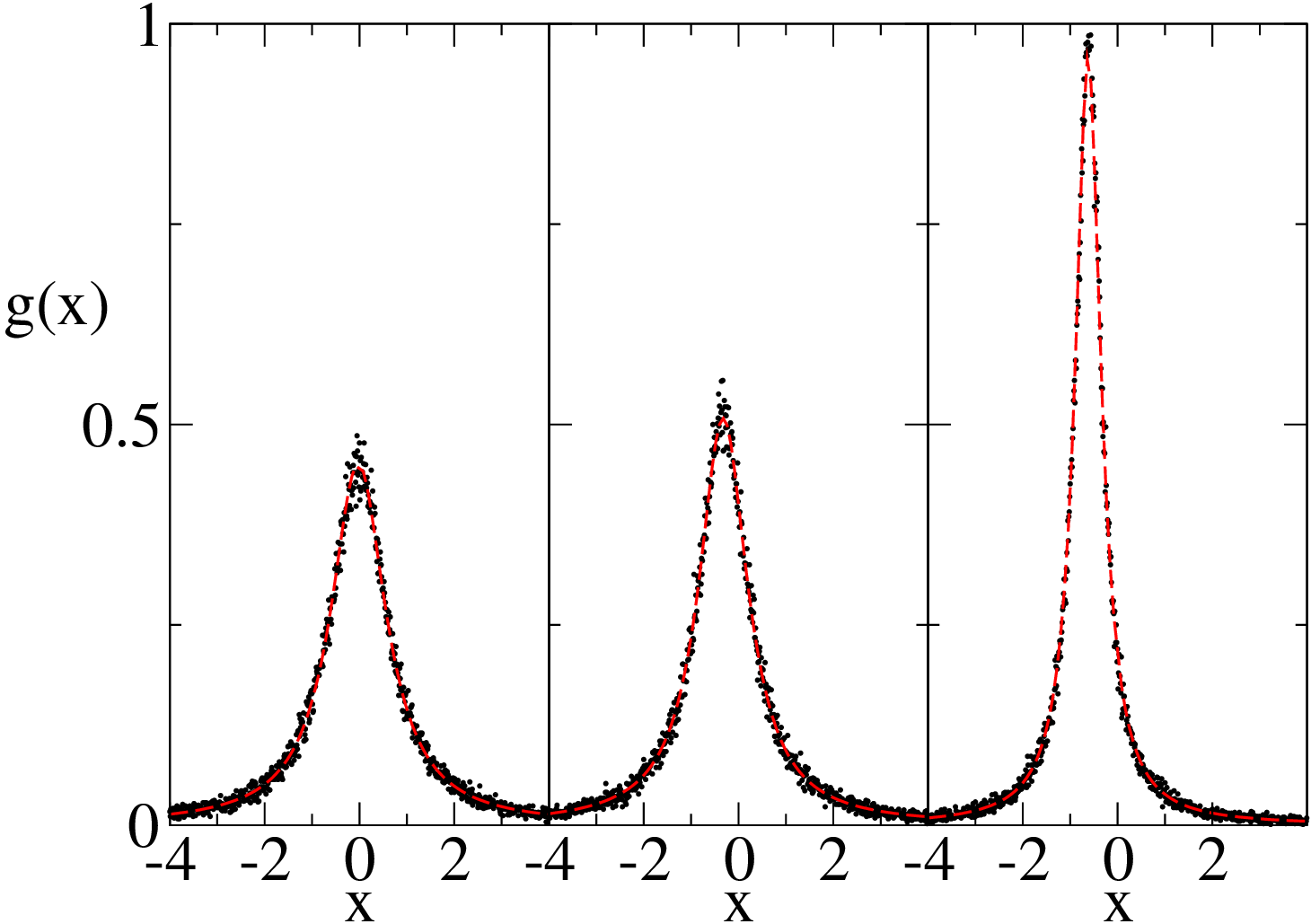}\\
\end{center}
\caption{Probability density $g(x)$ calculated numerically from the tree equation for regular random graphs with $K=2$ and $w=0.3$ at different energies, from left to right $E=0.1$, $1.3$ and $2.5$.  Black circles indicate results of direct numerical solution of the tree equation by belief propagation method, using a sample of $N=10^5$ initial values and performing $10^9$ iterations. Red dashed lines show the zeroth order (Cauchy) approximation \eqref{cauchy}. The next approximation \eqref{g_1} is practically indistinguishable from the zeroth one and is not shown on the plots.}
\label{w_0.3}
\end{figure}

\begin{figure}
\begin{center}
\includegraphics[width=.8\linewidth,clip]{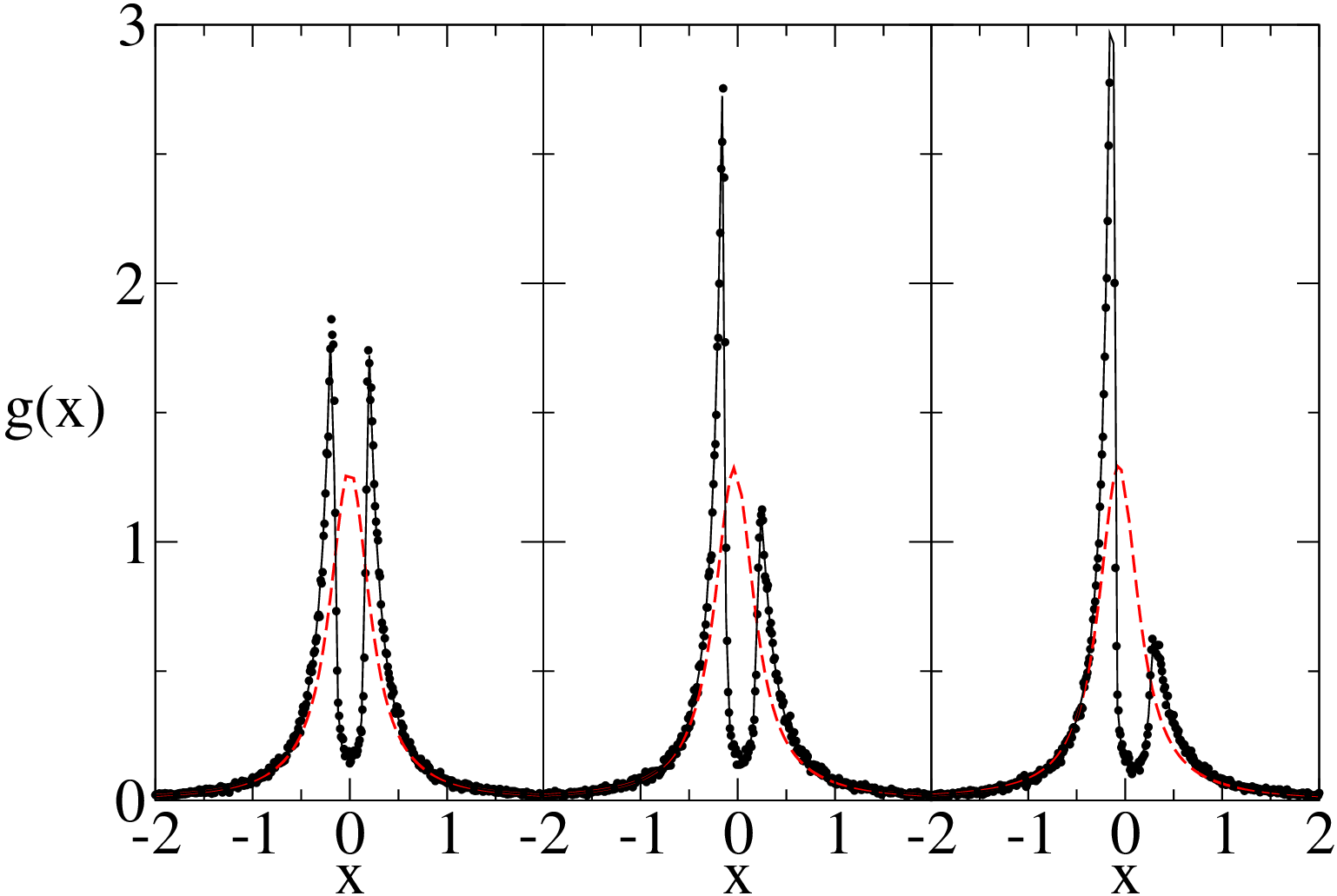}\\
\end{center}
\caption{Same as in Fig.~\ref{w_0.3} but with $w=12$. From left to right: $E=0.1$, $E=1.3$, and $E=2.5$. Black circles indicate results of direct numerical solution of the tree equation by belief propagation. Red dashed lines show the zeroth-order (Cauchy) approximation. Black solid lines are the first order  approximation \eqref{g_1}.  }
\label{w_12}
\end{figure}

To investigate more precisely  the accuracy of these different approximations, we calculate numerically the distribution $g(x)$ directly from the tree equation \eqref{te} by the belief propagation method explained in Section~\ref{introduction}. The results for various disorder strengths are presented in Figs.~\ref{w_0.3} and \ref{w_12}.  For each $w$ and each energy $E$, we calculate the zeroth order \eqref{cauchy} and the first order \eqref{g_1} approximations to $g(x)$. For small values of $w$ (Fig.~\ref{w_0.3}) the two approximations are practically the same but for larger $w$ they are clearly different. In all considered cases the first order approximation $g^{(1)}(x)$ given by Eq.~\eqref{g_1} is in a good qualitative agreement with direct numerical solution of the tree equation.

The first order approximation for the mean spectral density can be calculated e.g.~from Eq.~\eqref{modified_rho} which, for our model, reads
\begin{equation}
\rho^{(1)}(E)=\int_{-\infty}^{\infty} \frac{1}{z^2}g^{(1)}\Big (\frac{1}{z}\Big )g^{(1)}(z)\, \mathrm{d}z
\label{next_rho}
\end{equation} 
where $g^{(1)}(x)$ is calculated from \eqref{g_1}.
From Fig.~\ref{fig_diag_0.3} it is clear that this  approximation agrees better with direct numerical calculations at small $K$. For larger $K$ the zeroth order and the first order formulas are practically  indistinguishable in the scale of the figures. 
This robustness can be explained as follows.  As has been mentioned in Section~\ref{mean_field}, the choice of the Cauchy distribution as the zeroth order approximation is to a large degree arbitrary (what matters is only the asymptotics \eqref{limit}). On the other hand, according to the generalized central limit theorem the $K$-fold convolution, $F_K(x)$, of the chosen zeroth order function is in the bulk universal and is not sensitive to details of the initial function. Therefore, quantities which can be expressed through $F_K(x)$ and $F_{K+1}(x)$ at a low order are more precisely described by low-order approximations than those related directly with the initial function.
  

\section{Off-diagonal disorder}\label{off_diagonal}

In this Section we consider a different type of tree-like models, namely matrices associated with  regular graphs with fixed coordination number $K+1$, with diagonal elements set to zero and off-diagonal matrix elements defined as i.i.d.~random variables. The densities $g(x)$ and $\rho(E)$ are given by Eqs.~\eqref{general_g} and \eqref{general_rho} with $r(e)=\delta(e)$ (since diagonal elements are zero)  and a certain function $p(V)$ which determines  the probability density of off-diagonal elements. We choose as $p(V)$ the Gaussian distribution with variance $\sigma$ and zero mean
\begin{equation}
p(V)=\frac{1}{\sqrt{2\pi}\sigma}\mathrm{e}^{-V^2/(2\sigma^2)}.
\label{gaussian}
\end{equation}
This choice is not essential but simplifies analytical calculations. Note that changing the variance corresponds to a rescaling of the energy; therefore in what follows we set $\sigma=1$. 

It is convenient to define the integral
\begin{equation}
I_K(v)=\int_0^{\infty}\mathrm{e}^{-v t}\frac{1}{(1+t)^{K/2}}\, \mathrm{d}t,
\label{I_K}
\end{equation}
which for  integer $K$ reduces to well-known standard functions.  Many quantities defined in Section~\ref{mean_field} can be expressed through this integral. The transformations are straightforward and results are as follows. The 'mean-field' equation \eqref{ema} becomes 
\begin{equation}
\label{xtildeoffdiag2}
\tilde{x} =-\frac{1}{2\tilde{x}}I_K\left (\frac{E}{2\tilde{x}}\right )\ .
\end{equation}
The first order of the $K$-fold convolution, Eq.~\eqref{f_k_cauchy}, reads
\begin{equation}
F^{(1)}_K(z)=-\frac{1}{2\pi}\mathrm{Im} \left [ \tilde{x}^{-1}I_K\left (-\frac{z}{2\tilde{x}}\right )\right ]\ .
\end{equation}
The two approximations \eqref{mema} and \eqref{rho_sda} proposed in Section \ref{mean_field} for the spectral density are now given by
\begin{equation}
\rho_{\mathrm{MEMA}}=-\frac{1}{2\pi}\mathrm{Im} \left [ \tilde{x}^{-1}I_{1}\left (-\frac{1}{2\tilde{x}^2}\right )\right ]
\label{rho_mema}
\end{equation}
and 
\begin{equation}
\rho_{\mathrm{SDA}}=F_{K+1}(-E)=-\frac{1}{2\pi}\mathrm{Im} \left [ \tilde{x}^{-1}I_{K+1}\left (\frac{E}{2\tilde{x}}\right )\right ]\ . 
\label{sda}  
\end{equation}
The first order approximation of the distribution $g(x)$, obtained from the first iteration of the Cauchy function \eqref{g_1}, is readily expressed from Eq.~\eqref{general_g2} as
\begin{equation}
g^{(1)}(x)=\frac{1}{x^2}F^{(1)}_K(-E-\frac{1}{x})=-\frac{1}{2\pi x^2}\mathrm{Im} \left [ \tilde{x}^{-1}I_{K}\left (\frac{E+x^{-1}}{2\tilde{x}}\right )\right ]\ .  
\label{g_1_integral} 
\end{equation}
When $K\to\infty$ or $v\to\infty$, the main contribution in integral \eqref{I_K} comes from regions where $t\simeq 0$, and one can approximate $I_K(v)$ as
\begin{equation}
\label{Ikasympt}
I_K(v)\underset{K\to\infty}{\sim}\frac{1}{v+K/2}\ .
\end{equation}
The solution of Eq.~\eqref{xtildeoffdiag2} in this regime reads
\begin{equation}
\label{xtildeoffdiag}
\tilde{x} =\frac{-E}{2K}+\mathrm{i}\frac{\sqrt{4K-E^2}}{2K}.
\end{equation}
One can then check, using again \eqref{Ikasympt}, that in this limit $\rho_{\mathrm{MEMA}}$ and $\rho_{\mathrm{SDA}}$ given by \eqref{rho_mema} and \eqref{sda} tend to the Kesten-McKay law \eqref{mckay} for the regular graph
\begin{equation}
\rho^{(0)}(E)=\frac{K+1}{2\pi((K+1)^2\sigma^2-E^2)}\sqrt{4K\sigma^2-E^2},\qquad |E|\leq 2\sigma \sqrt{K}\ ,
\end{equation} 
where for convenience we reintroduce the variance $\sigma$ of off-diagonal elements. The limit  $K\to \infty$ gives the semi-circle density 
 \begin{equation}
\rho^{(0)}(E)=\frac{\sqrt{4K\sigma^2-E^2}}{2\pi K\sigma^2 },
\end{equation} 
as should be from general considerations \cite{theorem}.

For small values of $K$ the situation is different.   It is known (see e.g. \cite{off-diagonal}) that in graphs with off-diagonal disorder, where the probability of small values of $V$ is non-zero,  the mean density of states has a singularity at small energies of the form
\begin{equation}
\rho(E)\sim \rho(0)-\alpha |E|^{(K-1)/2},\qquad |E|\to 0\ ,
\label{peak}
\end{equation}
where $\alpha$ is a certain positive constant. This peak is related with a possibility of approximate localization at small sub-graphs which are isolated from the bulk due to small values of corresponding off-diagonal elements.   

\begin{figure}
\begin{center}
\includegraphics[width=.7\linewidth,clip]{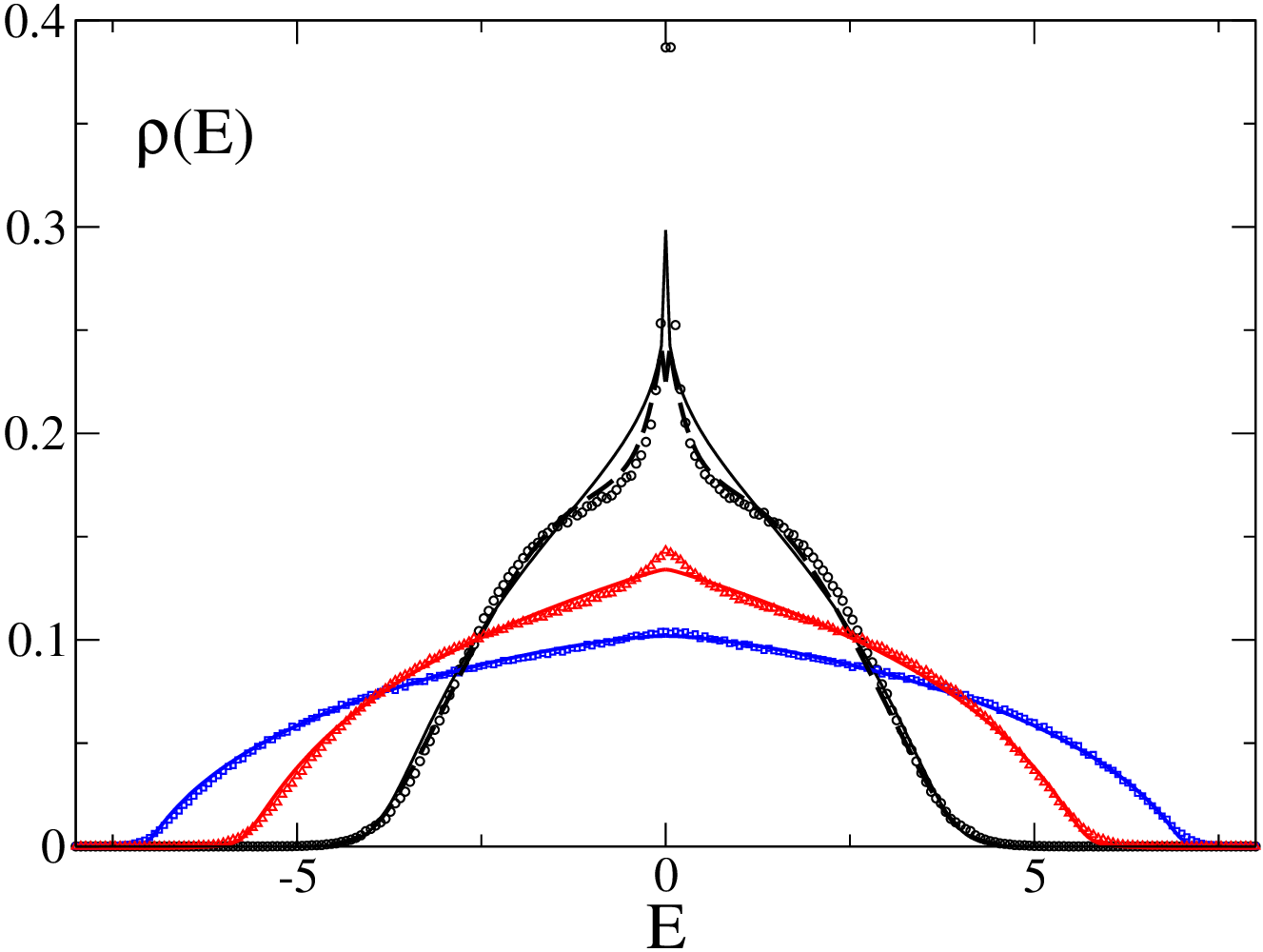}
\end{center}
\caption{Mean spectral density for regular random graphs with off-diagonal Gaussian disorder (with unit variance) for different values of the coordination number. Black circles correspond to $K=2$, red triangles to  $K=6$, and blue squares to $K=10$. Solid  lines of the same color  indicate the zeroth order MEMA approximations \eqref{rho_mema}. Dashed black line shows the next order approximation \eqref{double_integral} for $K=2$. For other values of $K$ the difference between the zeroth and the first order approximations is small and only the zeroth order approximation is indicated in the figure.  }
\label{off_fig}
\end{figure}

Numerical calculations for different $K$ and off-diagonal disorder given by \eqref{gaussian} are presented in Fig.~\ref{off_fig}. 
In the same figure  MEMA  expressions \eqref{rho_mema} are plotted. We checked that SDA formulas \eqref{sda} give worse  results at small $K$.  The overall agreement of approximate formulas and numerics is reasonably good but the presence of the peak \eqref{peak} is clearly visible especially at small $K$.  

\begin{figure}

\begin{minipage}{.49\linewidth}
\begin{center}
\includegraphics[width=.96\linewidth,clip]{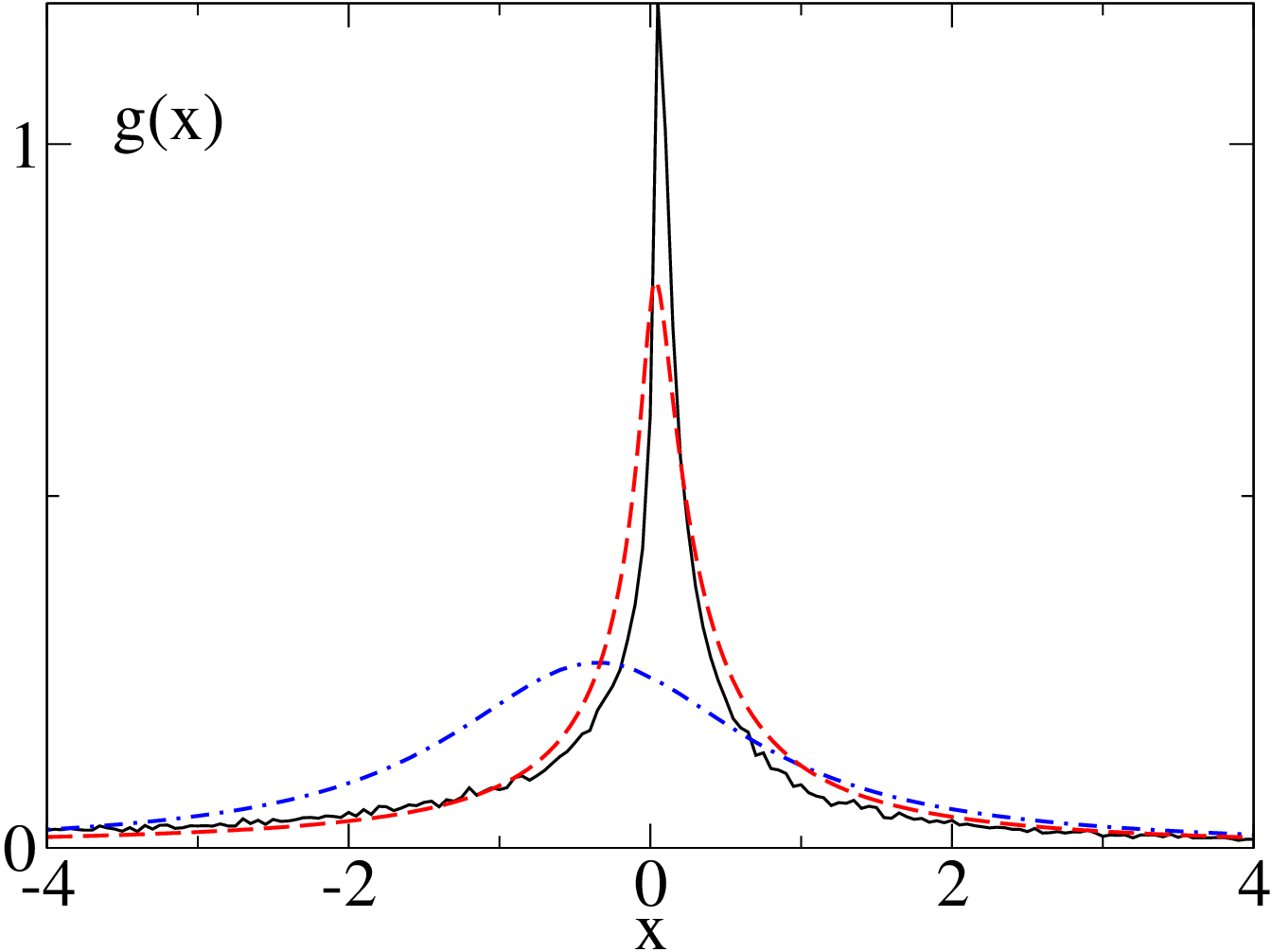}\\
(a)
\end{center}
\end{minipage}
\begin{minipage}{.49\linewidth}
\begin{center}
\includegraphics[width=.96\linewidth,clip]{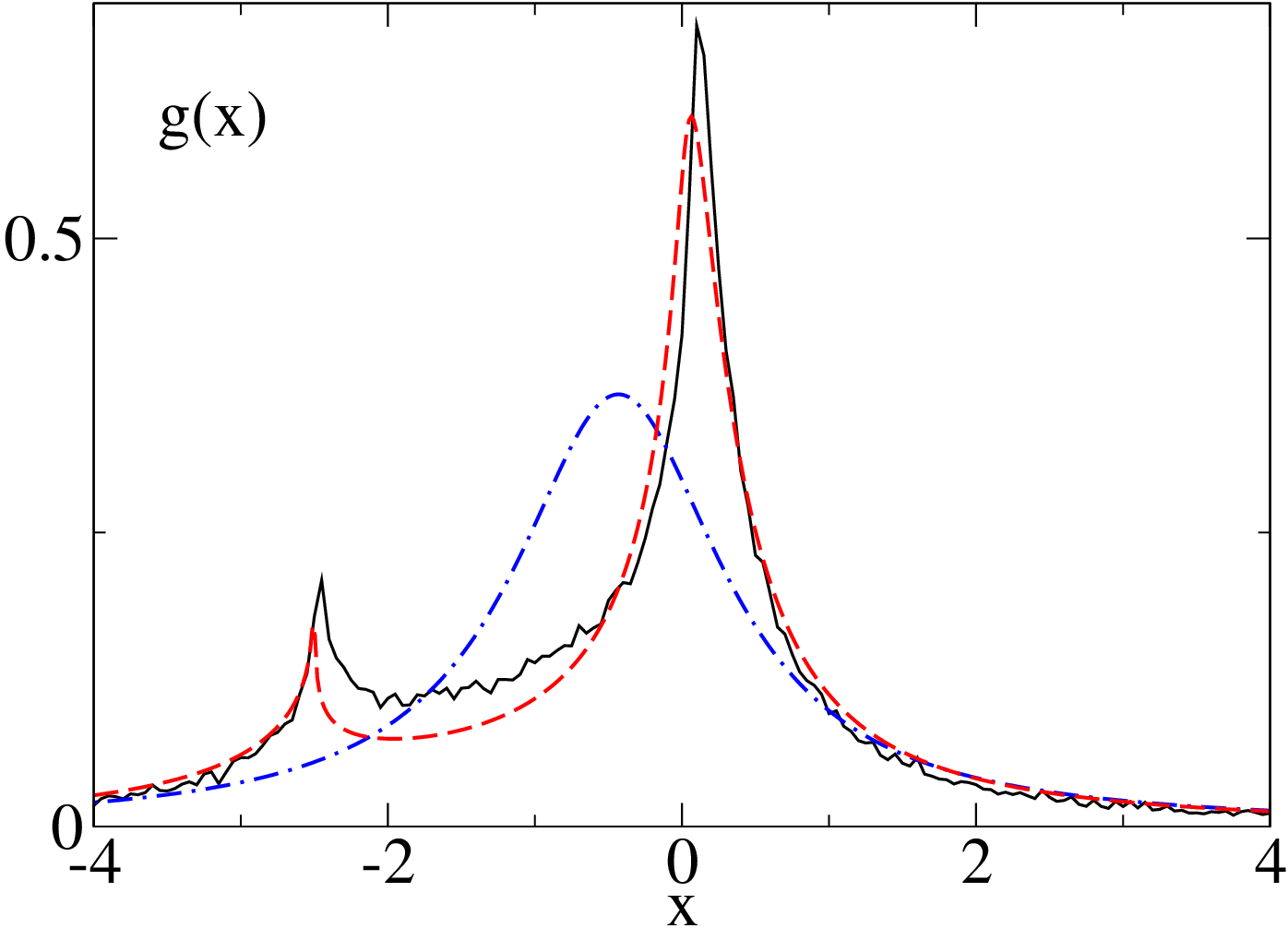}\\
(b)
\end{center}
\end{minipage}

\begin{minipage}{.49\linewidth}
\begin{center}
\includegraphics[width=.96\linewidth,clip ]{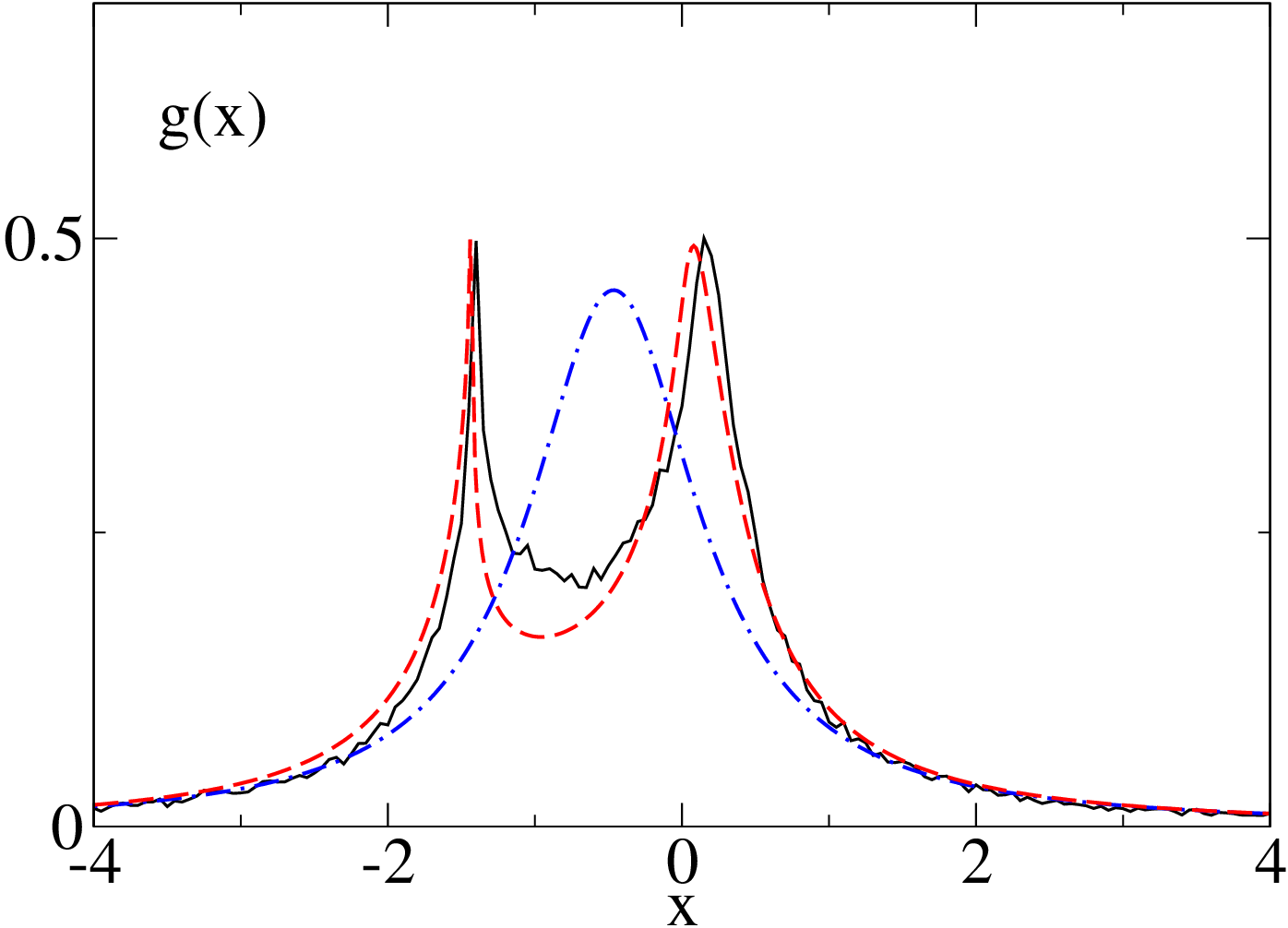}\\
(c)
\end{center}
\end{minipage}
\begin{minipage}{.49\linewidth}
\begin{center}
\includegraphics[width=.96\linewidth,clip]{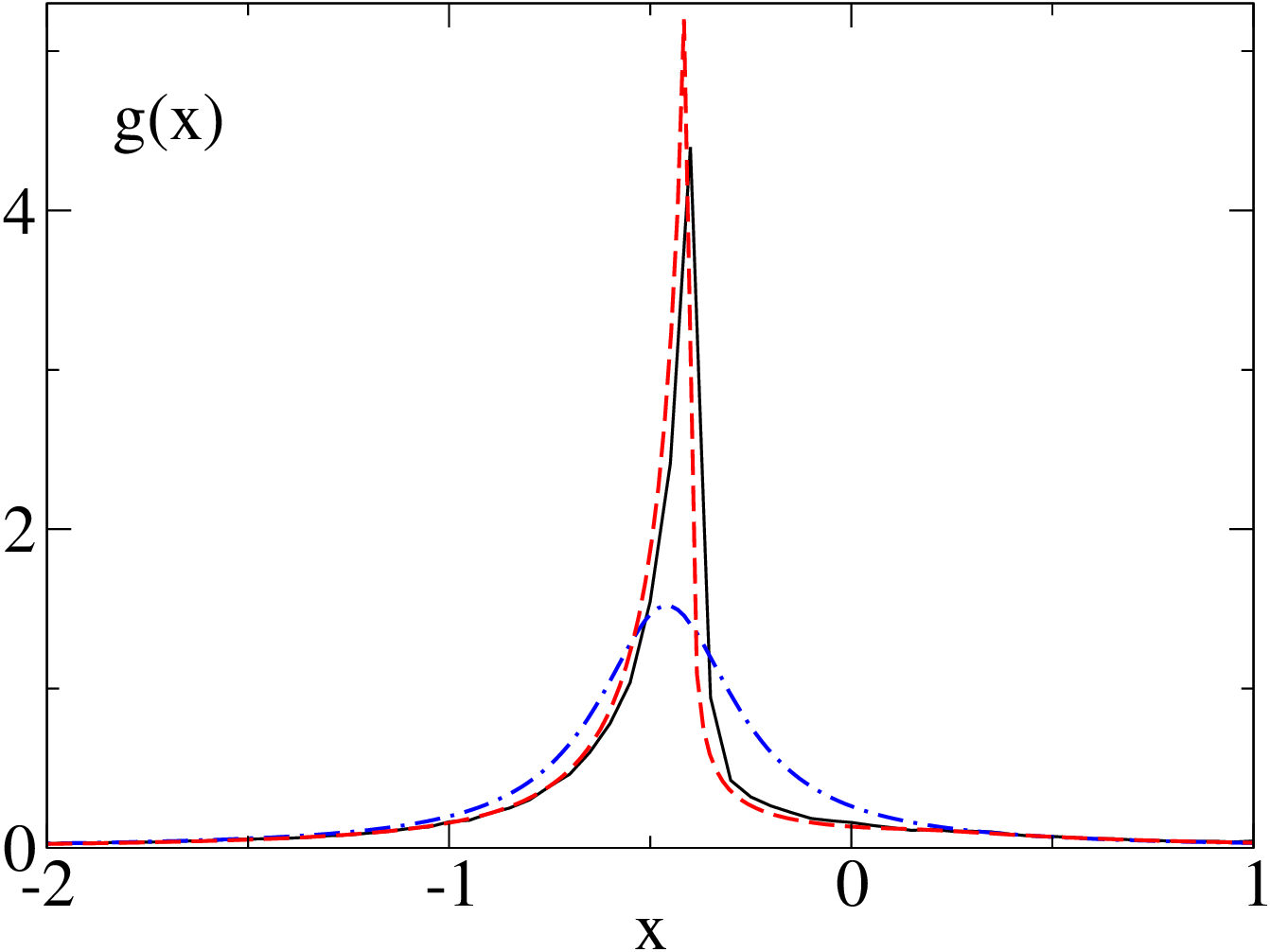}\\
(d)
\end{center}
\end{minipage}
\caption{Probability density $g(x)$ for regular graphs with  $K=2$ and off-diagonal disorder \eqref{gaussian}, at different energies: (a) $E=0.1$, (b) $E=0.4$, (c) $E=0.7$, (d) $E=2.5$. Solid black lines show results of direct numerical solution of the tree equation. Dashed-dotted blue lines indicate the zeroth order (Cauchy) approximation \eqref{cauchy} with parameters obtained from the 'mean-field' equation \eqref{xtildeoffdiag2}.   Dashed red lines are the first iteration of the above Cauchy distribution given by Eq.~\eqref{g_1}. }
\label{fig_functions}
\end{figure}

As in the previous Section we also calculate the function $g(x)$ by direct belief propagation method. The results are indicated in Fig.~\ref{fig_functions} for a few values of energy. The same figure also contains the zeroth-order (Cauchy) approximation \eqref{cauchy} and the first iteration of it given by \eqref{g_1_integral}. In all cases considered, the first order approximation is in a better agreement with numerical simulations. 

This better agreement of the first-order approximation for $g(x)$ motivated us to calculate the next approximation to the mean spectral density. We obtain it from \eqref{modified_rho} as
\begin{equation}
\label{rho1temp}
\rho^{(1)}(E)=\int \frac{1}{z^2 V^2 }g^{(1)}\Big (\frac{1}{z}\Big )g^{(1)}\Big (\frac{z}{V^2}\Big )p(V)\,  \mathrm{d}V\, \mathrm{d}z\ ,
\end{equation}
with $g^{(1)}(x)$ given by \eqref{g_1}. Introducing probability density $p_K(z)$ of $K$ i.i.d.~variables $V_j^2$ with distribution $p(V)$, 
\begin{equation}
p_K(z)=\int \delta(z-\sum_{j=1}^KV_j^2)\prod_{k=1}^K p(V_k)\, \mathrm{d}V_j,
\label{gaussian_convolution}
\end{equation}
one can rewrite \eqref{rho1temp} as
\begin{equation}
\label{rho1temp2}
\rho^{(1)}(E)=\int \frac{1}{tz^2}g_{\chi_1}\left(\frac{1}{z}\right)g_{\chi_2}\left(\frac{z}{t}\right)p_1(t)p_K(t_1)p_K(t_2)\, \mathrm{d}z\,  \mathrm{d}t\, \mathrm{d}t_1\, \mathrm{d}t_2\ , 
\end{equation}
with
\begin{equation}
\chi_i=-\frac{1}{E+\tilde{x}t_i}\ ,\qquad i=1,2
\end{equation}
(note that since $t_i\leq 0$ and Im$(\tilde{x})>0$ one has Im$(\chi_i)>0$). Using the elementary identities \eqref{idcauchy1}--\eqref{idcauchy2} for the Cauchy distribution, one directly gets
\begin{equation}
\rho^{(1)}(E)=-\frac{1}{\pi}\int \mathrm{Im}\, \left (\dfrac{1}{E+\tilde{x}t_2-\dfrac{t}{E+\tilde{x}t_1}} \right ) p_1(t) p_K(t_1)p_K(t_2)\,  \mathrm{d}t\, \mathrm{d}t_1\, \mathrm{d}t_2\ .
\label{off_rho_second}
\end{equation}
For the Gaussian disorder \eqref{gaussian} with $\sigma=1$,
\begin{equation}
p_K(z)=\frac{1}{2^{K/2}\Gamma(K/2)}z^{(K-2)/2}\mathrm{e}^{-z/2}\ 
\end{equation}
for $z>0$, zero otherwise. The triple integral in \eqref{off_rho_second} are simplified by the transformation
\begin{equation}
\rho^{(1)}(E)=\frac{1}{\pi}\int_0^{\infty} \,\mathrm{d} \beta \int \mathrm{Re}\,\exp \left [\mathrm{i}\beta \Big ( E+\tilde{x}t_2-\frac{t}{E+\tilde{x}t_1} \Big ) \right ]p_1(t) p_K(t_1)p_K(t_2) \mathrm{d}t\, \mathrm{d}t_1\, \mathrm{d}t_2 
\end{equation}
together with  the integration over $t$ and $t_2$ using \eqref{gaussian_convolution}. It leads  (after changing $\beta\to\mathrm{i}\beta$) to 
\begin{equation}
\rho^{(1)}(E)=-\frac{1}{ 2^{K/2}\pi \Gamma(K/2) } \mathrm{Im}\, \int_0^{\infty} t^{(K-2)/2}\mathrm{e}^{-t/2}\,  \mathrm{d}t \int_0^{\infty } \dfrac{\mathrm{e}^{-\beta E}}{(1+2 \beta \tilde{x})^{K/2}\sqrt{1-\dfrac{2\beta}{E+\tilde{x}t}}}\,  \mathrm{d}\beta\ .
\label{double_integral}
\end{equation}
This expression is plotted in Fig.~\ref{off_fig} for $K=2$. It clearly agrees better with the direct numerical calculation of the mean spectral density. For larger values of $K$ the zeroth and the first order formulas are close to each other. It confirms the statement observed in the previous Section that though the first order approximation  much better agrees with probability density $g(x)$ obtained by the belief propagation, the difference between the two approximations in the mean density is noticeable only for the smallest values of coordination number.    
   
In the same way it is possible to investigate the general case of regular graphs with both diagonal and off-diagonal disorders. We mention only a special case of matrices of the following form
\begin{equation}
M_{mn}=e_n\delta_{mn}+ \frac{1}{\sqrt{N}}V_{mn}, \qquad m,n=1,\ldots, N
\label{pastur_model}
\end{equation}
where $e_n$ are i.i.d.~random variables with probability density $r(e)$ and $V_{mn}$ are i.i.d.~real symmetric variables with zero mean and finite variance
\begin{equation}
\langle V_{mn}\rangle =0,\qquad \langle V_{mn}^2\rangle =V^2.
\end{equation}
Pastur proved in \cite{pastur} that in the limit $N\to\infty$ and under certain mild conditions the mean Green function 
\begin{equation}
G(E)=\frac{1}{N}\sum_{n}\langle G_{nn}\rangle 
\label{G_E}
\end{equation}
obeys the equation
\begin{equation}
G(E)=\int \frac{r(e)\mathrm{d}e}{e-E-V^2\ G(E)}\ .
\label{pastur_equation}
\end{equation} 
In the formalism discussed here this case corresponds to a regular graph with coordination number $K=N-1$. As in the limit $N\to\infty$ $G(E)$ in \eqref{G_E} tends to  $\tilde{x}(E)$ (cf. Eq.~\eqref{rho_ema}) and since from the definition \eqref{pastur_model} 
\begin{equation}
\lim_{N\to\infty}\sum_{n\neq m}M_{mn}^2=V^2,
\end{equation}
it is easy to check that Eq.~\eqref{pastur_equation} coincides with the 'mean-field' equation \eqref{ema}, which gives another confirmation of the  tree-equation universality.
\section{Conclusion}\label{conclusion}

We investigated the mean spectral density for random regular graphs by finding an approximate real solution of the tree equation associated with these graphs. This equation is general and is valid for any uniform tree-like models. The mean spectral density for these models is determined by the $(K+1)$-fold convolution of real tree equation solution, where $K+1$ is the coordination number of the tree. For large $K$ the generalized central limit theorem states that in the bulk such convolution depends only on a few parameters which can be calculated directly from the initial function.  From the structure of the tree equation it follows that the required solution belongs to the domain of attraction of the symmetric Cauchy distribution. Therefore, it is natural to use, as the zeroth order approximation of the tree equation solution,  the Cauchy distribution itself, whose parameters are calculated self-consistently. Iterations of that initial Cauchy distribution give next order approximations. When a good approximation for tree equation solution is found, the mean spectral density can be calculated using one of the exact formulas relating it to the tree equation solution (see Section~\ref{introduction}). 

We applied this scheme for the calculation of mean spectral density for regular graphs with diagonal or off-diagonal disorder, and compared the zeroth and first order approximations with results of direct numerical calculations. As expected, for large coordination number the zeroth order approximation for the density gives quite good results, which are rather accurate even at the smallest $K$. For regular graphs with diagonal disorder the SDA \eqref{rho_sda} gives a slightly better results but for graphs with off-diagonal disorder the MEMA \eqref{mema} is closer to the numerics than other zeroth order approximations. 

As for the tree equation solution itself, the first order approximation is always much closer to the numerical solution obtained by the belief propagation than the zeroth order Cauchy approximation. Nevertheless, the corresponding discrepancies for the mean spectral density are usually noticeable only at small coordination numbers.  

The statement that the mean spectral density of random graphs is related to the solution of an equation which, strictly speaking, is valid only for corresponding trees is physically quite natural and could be proved rigorously in certain cases without disorder and with diagonal disorder. Further investigation of this and related questions in the spirit of trace formulas on graphs (see e.g. \cite{uzy}) is of interest. 
      
\begin{acknowledgments}
The authors thank Y. Fyodorov for pointing out Ref.~\cite{yan}, and D. Jakobson for bringing the attention to Ref. \cite{geisinger}.
\end{acknowledgments}


\end{document}